\documentclass[final,12pt,3p]{elsarticle}

\usepackage{graphicx}
\usepackage{hyperref}
\hypersetup{
backref=true, 
bookmarks=true, 
unicode=false, 
pdftoolbar=true, 
pdfmenubar=true, 
pdffitwindow=false, 
pdfstartview={FitH}, 
pdftitle={}, 
pdfauthor={}, 
pdfsubject={}, 
pdfkeywords={}{}, 
pdfnewwindow=true, 
colorlinks=true, 
linkcolor=BlueViolet, 
citecolor=maroon, 
urlcolor=blue,
}
\usepackage[utf8]{inputenc}
\usepackage{subfig}
\usepackage{caption}

\usepackage{amssymb}
\usepackage{gensymb}

\usepackage{natbib}
\biboptions{comma,round}
\setcitestyle{authoryear}

\journal{Climate Change Economics}

\begin{document}

\begin{frontmatter}

\title{Costs and Benefits of the Paris Climate Targets}

\author[label1,label2,label3,label4,label5,label6]{Richard S.J. Tol\corref{cor1}}
\address[label1]{Department of Economics, University of Sussex, Falmer, United Kingdom}
\address[label2]{Institute for Environmental Studies, Vrije Universiteit, Amsterdam, The Netherlands}
\address[label3]{Department of Spatial Economics, Vrije Universiteit, Amsterdam, The Netherlands}
\address[label4]{Tinbergen Institute, Amsterdam, The Netherlands}
\address[label5]{CESifo, Munich, Germany}
\address[label6]{Payne Institute for Public Policy, Colorado School of Mines, Golden, CO, USA}

\cortext[cor1]{Jubilee Building, BN1 9SL, UK}

\ead{r.tol@sussex.ac.uk}
\ead[url]{https://research.vu.nl/en/persons/rsj-tol}

\begin{abstract}
The temperature targets in the Paris Agreement cannot be met without very rapid reduction of greenhouse gas emissions and removal of carbon dioxide from the atmosphere. The latter requires large, perhaps prohibitively large subsidies. The central estimate of the costs of climate policy, unrealistically assuming least-cost implementation, is 3.8-5.6\% of GDP in 2100. The central estimate of the benefits of climate policy, unrealistically assuming constant vulnerability, is 2.8-3.2\% of GDP. The uncertainty about the benefits is larger than the uncertainty about the costs. The Paris targets do not pass the cost-benefit test unless risk aversion is high and discount rate low.
\end{abstract}

\begin{keyword}
climate policy \sep net zero \sep cost-benefit analysis
\JEL Q54
\end{keyword}

\end{frontmatter}

\section{Introduction}
International targets for climate policy are political. The upper limit of the temperature target of the 2015 Paris Agreement under the United Nations Framework Convention on Climate Change (UNFCCC) can be traced back to an old and flawed report by an advisory council \citep[][cf. \citet{Tol2007EP}]{WBGU1995}, but the lower limit cannot even claim dubious support. Some of the countries that have adopted the Paris Agreement require cost-benefit analysis of policy decisions, but this requirement does not extend to international treaties. The Intergovernmental Panel on Climate Change (IPCC) does not offer cost-benefit analysis either, indeed has shied away from reviewing the relevant academic literature on this matter \citep{Tol2022ipcc}. This paper therefore reports a cost-benefit analysis of the targets in the Paris climate agreement.

This is surely not the first cost-benefit analysis of climate policy. That honour goes to \citet{Nordhaus1982}. Since then, there have been many attempts to balance the costs and benefits of greenhouse gas emission reduction \citep[e.g.,][]{Nordhaus1992, Peck1994, Tol1999kyoto, Keller2004, Tol2012EP, Millner2013, Crost2014, Barrage2020, Bremer2021}. All of these papers support greenhouse gas emission reduction but few \citep{Hansel2020} advocate 100\% reduction of carbon dioxide emissions, which is needed to stabilize its atmospheric concentration and so halt anthropogenic climate change. The reason lies in the structure of cost-benefit analysis, which equates the marginal costs of emission reduction to its marginal benefits. Cost-benefit analysis rarely recommends a corner solution. Although the emergence of negative carbon energy and direct air capture imply that a 100\% emission reduction is not a corner solution in cost, it is in benefit: If climate would not change any more, the impact of climate change would be zero, and the marginal impact of emissions would be close to zero. Stringent emission reduction thus reduces and eventually removes the justification for even more stringent emission reduction.

There are two exceptions to this. First, some modellers assume a backstop technology \citep{Nordhaus2014}, which, at a finite carbon tax, would fully and irreversibly decarbonize the economy, even if that carbon tax is subsequently withdrawn. This is a strong assumption. The other exception is the assumption that climate rather climate \emph{change} damages the economy \citep{Burke2015}, in which cases any stable climate that is warmer than pre-industrial would have marginal damages substantially above zero. This assumption too is hard to support \citep{Newell2021}.

This paper contributes a cost-benefit analysis of the two temperature targets of the Paris Agreement. Instead of finding the optimal temperature, I assess whether these targets pass the cost-benefit test. I do this on the basis of (\textit{i}) the latest IPCC estimates of the costs of emission reduction \citep{Rogelj2018IPCC, Riahi2022IPCC} and (\textit{ii}) a new meta-analysis of the impact of climate change \citep{Tol2009JEP, Howard2017, NordhausMoffat2017}.

The paper proceeds as follows. Section \ref{sc:targets} assesses just how ambitious the Paris targets are. Section \ref{sc:costs} reviews the costs of climate policy and Section \ref{sc:benefits} its benefits. Section \ref{sc:cba} reports a cursory cost-benefit analysis. Section \ref{sc:conclude} concludes.

\section{The scale of ambition}
\label{sc:targets}
\begin{figure}[h]
    \centering
    \includegraphics[width=\textwidth]{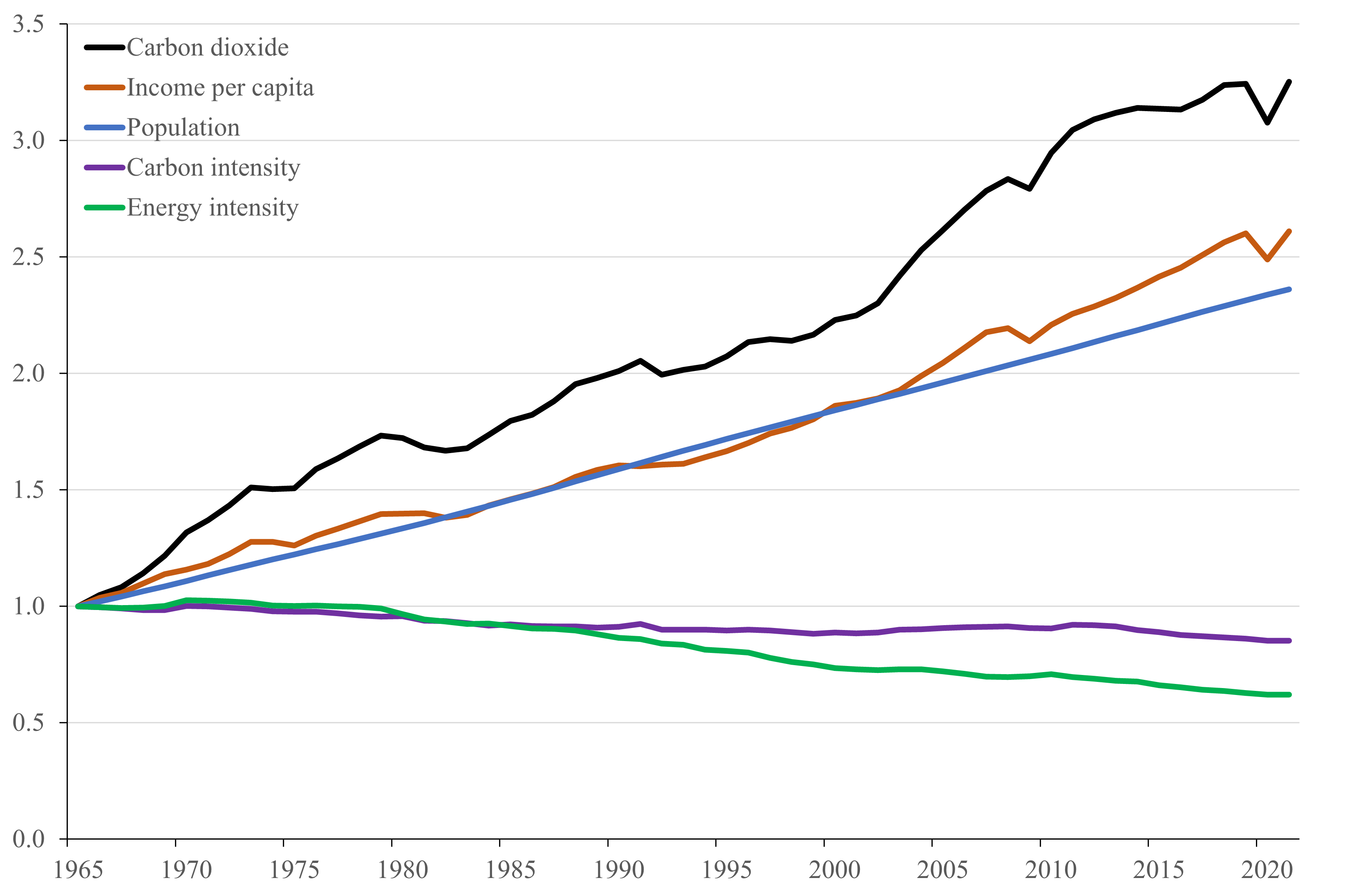}
    \caption{Carbon dioxide emissions from energy use, 1965-2021, and the four components of the Kaya Identity}
    \label{fig:kaya}
\end{figure}

Figure \ref{fig:kaya} shows global carbon dioxide emissions from fossil fuel combustion for the period 1965-2021, the period for which we have good data. Emissions rose by 2.1\% per year on average, but growth slowed to 0.7\% in the most recent decade. An annual emission reduction of 15\% for 28 years would reduce global emissions to close to zero. Global net zero emissions by 2050 is needed to meet the 1.5\celsius{} Paris target.

Note that China aims at net zero emissions by 2060, and India by 2070. Global net zero by 2050 therefore means net negative emissions in the OECD.

Figure \ref{fig:kaya} also shows the components of the Kaya Identity. Population growth was 1.5\% per year over the full period but slowed to 1.1\% in the last decade. See Table \ref{tab:kaya}. Per capita income grew by an annual 1.7\% between 1965 and 2021; this slowed to 1.5\% after 2011. Assuming that these two components are largely beyond the remit of climate policy, emission reduction has to come from improvements in energy efficiency and carbon intensity. Over the full period, energy intensity, the amount of primary energy needed to generate one dollar of value added, fell by 0.9\% per year, accelerating to 1.1\% in the last ten years. Carbon intensity, the amount of carbon dioxide emitted per primary energy unit used, fell by 0.3\% per year over the whole period and by 0.8\% in the last decade.

If emissions are to fall by 15\% per year while the economy continues to grow by 2.5\% per year, the sum of energy and carbon intensity has to fall by 17.5\% per year, up from 1.9\% in the last ten years, the period of most intense climate policy. At first sight, the scale of ambition in international climate policy is momentous.

Renewables are one of the drivers of the slower growth of emissions. Integrating non-dispatchable electricity becomes more expensive as its share in power generation grows. Electricity is probably the easiest sector to decarbonize. It is more difficult for transport, heating, industry, and agriculture. That is, an order of magnitude increase in the decarbonization rate requires more, much more than a tenfold increase in the policy effort. The low-hanging fruit has already been picked.

Furthermore, the energy sector is characterised by long-lived capital. A lot of the buildings, power plants, steel mills and chemical plants we use today will still be around in 2050, and even some of the machinery and vehicles \citep{Davis2010, Tong2019}. That is why the target is \emph{net} zero. \emph{Gross} zero would require capital destruction at a large scale, with bankruptcies, lay-offs, and claims for compensation. Net zero emission requires afforestation\textemdash large plantations of rapidly growing trees\textemdash negative carbon energy\textemdash electricity generated from biomass with carbon capture or storage\textemdash and direct air capture\textemdash removing carbon dioxide with artificial photosynthesis. Scale and speed are the problems with afforestation. Agricultural lands are already converting back to nature in Europe and North America. This can be accelerated but not by much, and only at the expense of diverse forests, including slow-growing species. Scale is also the problem with bioenergy. Cheap biofuel requires large, heavily mechanized monoplantations. The acreage needed to supply the required energy is infeasibly large \citep{Wise2009}. Direct air capture is a proven technology at a small scale \citep{House2011}; both scaling up and safe storage of large volumes of carbon dioxide are problematic.

\section{The costs of emission reduction}
\label{sc:costs}
Emission reduction costs money \citep{Weyant1993, Clarke2014}. Models agree that a complete decarbonization of the economy can be achieved at a reasonable cost if policies are smart, comprehensive and gradual and if targets are sensible. Models disagree on how much emission reduction would cost; estimates vary by an order of magnitude or more. \citet{Riahi2022IPCC} reports that the global average carbon tax needed to meet the 1.5\celsius{} temperature target ranges between \$30/tCO\textsubscript{2} and \$1,100/tCO\textsubscript{2} in 2030 and between \$110/tCO\textsubscript{2} and \$14,000/tCO\textsubscript{2} in 2100. That target would cost somewhere between 0.5\% and 6.0\% of GDP in 2030, and up to 10\% in 2100. 

\citet{Barker2007} and \citet{Clarke2009} found that the 2\celsius{} target is infeasible for physical, technical, economic or political reasons. Modellers have met the political demand for more stringent targets by expanding options for negative emissions \citep{Clarke2014, Riahi2022IPCC}. As the market for carbon dioxide is typically saturated, negative emissions require a carbon subsidy (and deserve one, as this is a negative externality). \citet{Tol2019HB} finds that the central estimate of these subsidies amount to 4\% of world income by the end of the century, with one model putting it at almost 17\%. These are global averages\textemdash net positive emissions in Asia after 2050 would have to be offset by net negative emissions in Europe and North America.

The cost estimates cited above typically assume cost-effective implementation of climate policy. Under ideal conditions, first-best regulation is straightforward: The costs of emission reduction should be equated, at the margin, for all sources of emissions \citep{Baumol1971}. Governments routinely violate this principle, with different implicit and even explicit carbon prices for different sectors and for differently sized companies within sectors. Although climate change is a single externality, emitters are often subject to multiple regulations on their greenhouse gas emissions \citep{Boehringer2008, Boehringer2010}. Regulations are often aimed at a poor proxy for emissions (e.g., car ownership) rather than at emissions directly \citep{Proost2001}, and instrument choice may be suboptimal \citep{Webster2010}. Conditions are not ideal. Optimal policy deviates from the principle of equal marginal costs to accommodate for market power \citep{Buchanan1969}, for multiple externalities \citep{Ruebbelke2003, Parry2005}, and for prior tax distortions \citep{Babiker2003}. Such deviations are subtle and context-specific, and rarely observed in actual policy design. All this makes that \emph{actual} climate policy is far more expensive than what is assumed in models.

While the above problems are well-known to those who know it well, there is another issue that has attracted little or no attention. The cost of greenhouse gas emission reduction is typically reported as a drop in GDP. Although GDP is not a welfare measure, this is fine as first sight as Gross Domestic Product is theoretically equal to Gross Domestic Income, and income is roughly proportional to consumption, which is in turn a closely related to utility. However, climate policy not only changes the size of GDP but also its composition. Particularly, stringent temperature targets require removing carbon dioxide from the atmosphere. This is an economic activity and so contributes to GDP. As a defensive expenditure that does only serves to prevent welfare loss, carbon dioxide removal does not count towards the Indicator of Sustainable Economic Welfare \citep{NordhausTobin1972} or the Environmental Net Product \citep{Hartwick1990}. The number cited above, 4\% of GDP, is based on \emph{net} negative emissions; assuming a continued use of fossil fuels, \emph{gross} negative emissions and so subsidies / defensive expenditures would be larger.

\begin{figure}[h]
    \centering
    \includegraphics[width=\textwidth]{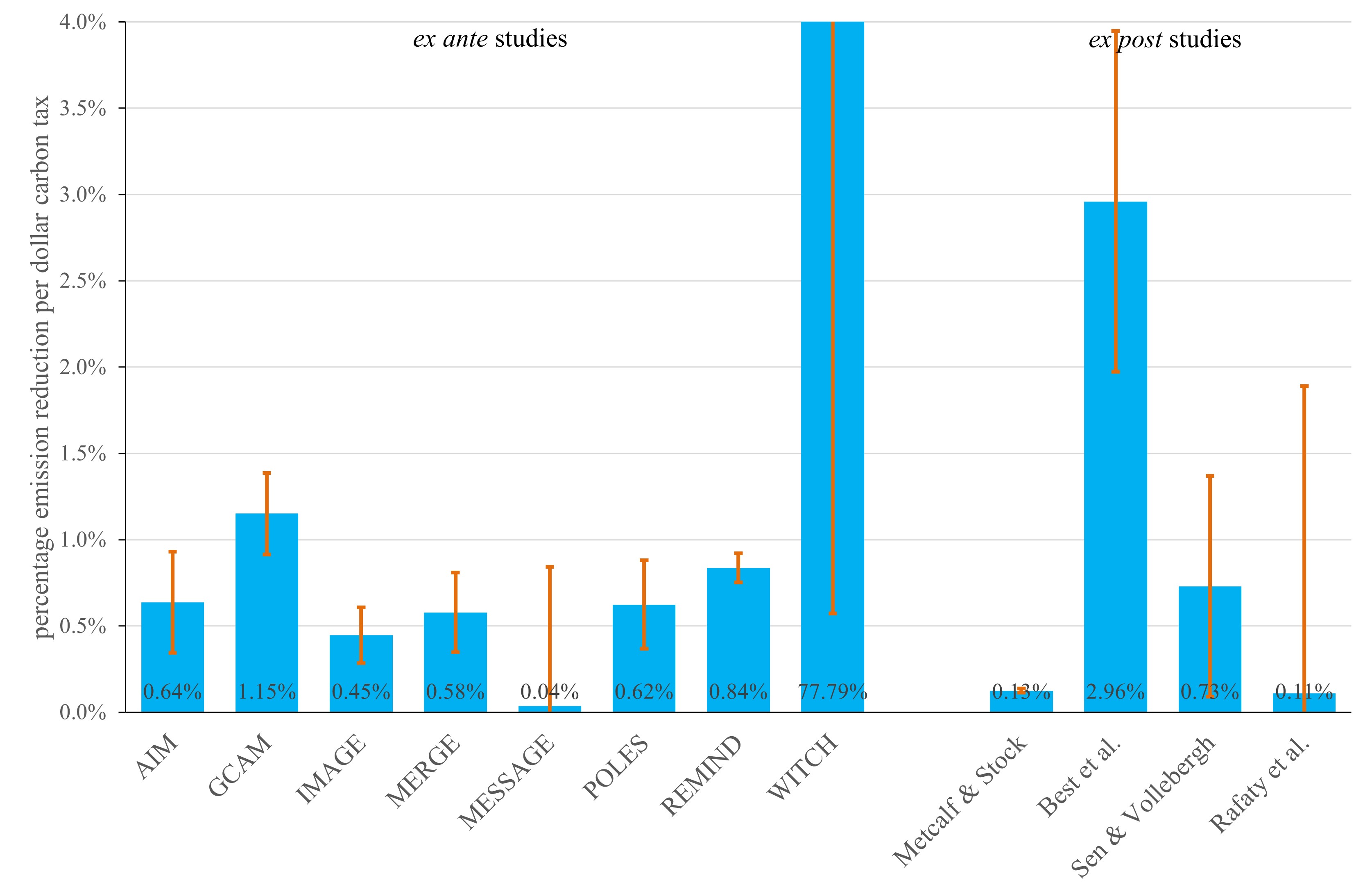}
    \caption{The efficacy of a carbon tax according to eight \textit{ex ante} and four \textit{ex post} studies.}
    \label{fig:taxeff}
\end{figure}

Unfortunately, the AR6 scenario database has yet to be released. The scenario database for the IPCC Special Report on 1.5\celsius{} \citep{Rogelj2018IPCC} is available, allowing for a closer inspection of results than the graphical summaries in IPCC reports.\footnote{Our World in Data offers a \href{https://ourworldindata.org/explorers/ipcc-scenarios}{Scenario Explorer}.} Figure \ref{fig:taxeff} shows the efficacy of carbon pricing according to eight different models. Tax efficacy is here defined as the reduction in carbon dioxide emissions in 2030 divided by the assumed carbon prices in the 2020s. Tax efficacy varies over two orders of magnitude, from 0.04\% for \textsc{message} to 1.15\% tCO\textsubscript{2}/\$ for \textsc{gcam}.\footnote{\textsc{witch} has perfect foresight, responds very strongly to future carbon prices.}

Figure \ref{fig:taxeff} also shows tax efficacy as measured by four \textit{ex post} studies, that econometrically estimate the impact of carbon taxes on emissions \citep[][see \citet{Tol2022ipcc} for a discussion of \textit{ex post} studies for a broader set of climate policies.]{Sen2018, Best2020, Metcalf2020, Rafaty2020}. The results are diverse. Three studies find significant effects, but one does not. One study is well in line with 6 of the 8 \textit{ex ante} studies, one finds larger tax efficacy, and two find a lower tax efficacy. Comparison between \textit{ex ante} and \textit{ex post} estimates is not one-on-one, as the former assume first-best policy implementation. For example, the carbon taxes studied by Metcalf and Stock are imposed on hard-to-abate sectors outside the EU ETS. The two sets of studies agree, however, that current estimates are not exactly firm.

\section{The benefits of emission reduction}
\label{sc:benefits}
\citet{Tol2022meta} reviews 39 papers with 61 published estimates of the total economic impact of climate change.

\begin{figure}[h]
\includegraphics[width=\textwidth]{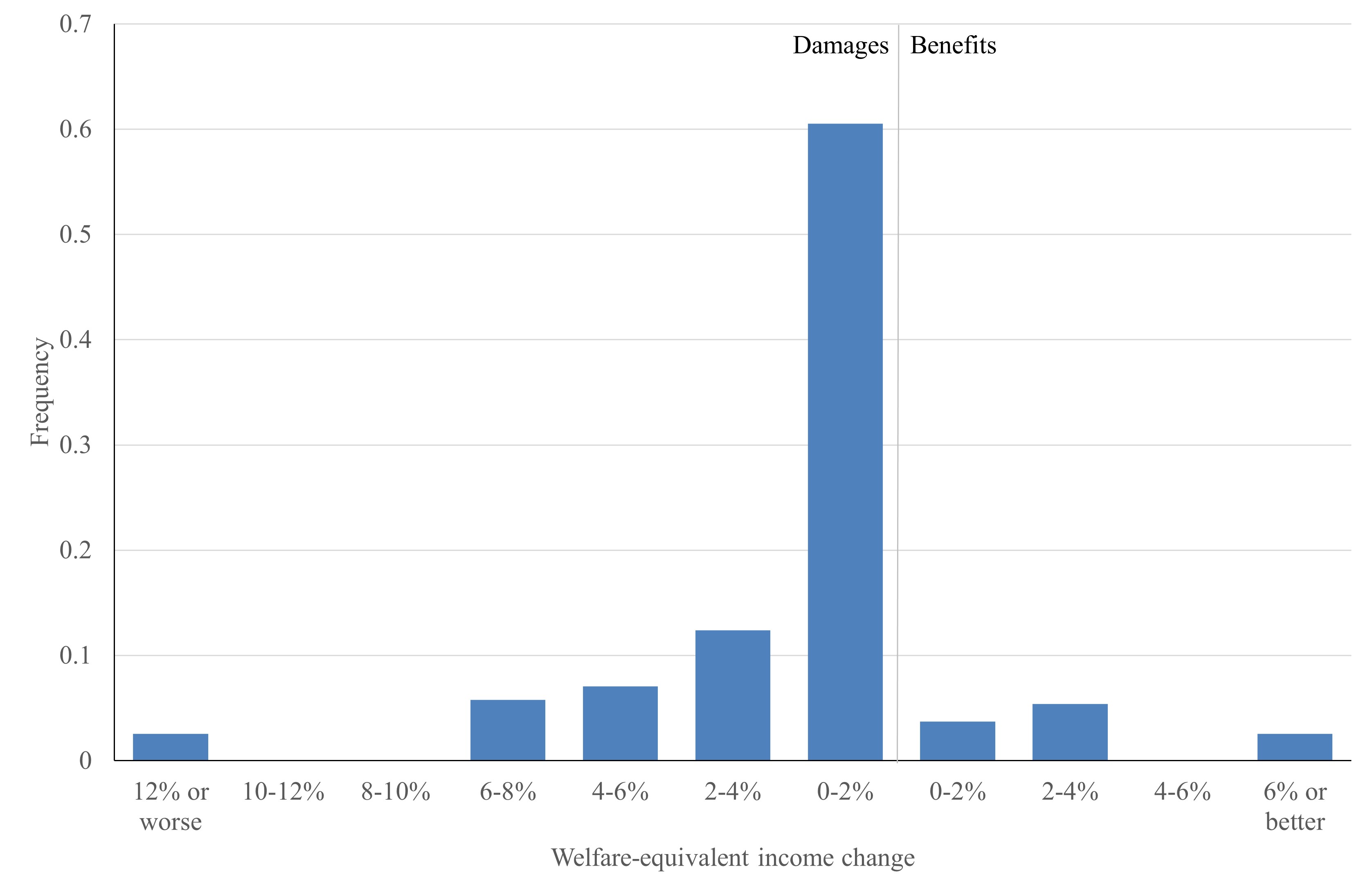}
\caption{Histogram of total economic impact of 2.2\celsius{} global warming as estimated.}
\label{fig:totalimpact}
\end{figure}

Figure \ref{fig:totalimpact} shows the histogram of published estimates. Estimates are published for a range of climate change scenario. \citet{Tol2022meta} fits seven alternative impact functions to these estimates and uses the fit as weights in the weighted average impact function. This model average is used here to scale all estimates to 2.5\celsius{} warming. The histogram includes all 61 estimates, but weighted such that each of the 39 papers contributes 1/39 to the total frequency.

Some 60\% of estimates show moderate damages, between 0 and 2\% of GDP \citep{Pearce1996IPCC, Arent2014}. The central estimate of the welfare change caused by a century of climate change is comparable to the welfare loss caused by losing a year of economic growth.

\citet{Tol2022meta} finds that these \textit{ex ante} estimates are not inconsistent with \textit{ex post} econometric studies of the impact of weather shocks on economic growth\textemdash for those studies that relate economic growth to temperature change. Econometric studies that relate economic growth to temperature levels show much larger impacts, positive or negative, but suffer from both econometric problems \citep{Newell2021} and conceptual ones\textemdash notably the implication that climate change would have a permanent effect on economic growth, a form of climate determinism that contradicts all empirical evidence.

The uncertainty about the central estimates is rather large, however, and benefits cannot be excluded, even for high warming. About 12\% of estimates show benefits rather than damages. These benefits are due to reduced costs of heating in winter, reduced cold-related mortality and morbidity, and carbon dioxide fertilization, which makes plants grow faster and more resistant to drought. Negative impacts, such as summer cooling costs, infectious diseases, and sea level rise, dominate the central estimate.

The uncertainty about the welfare impact of climate change is not just large, it is also right-skewed. Around 38\% of estimates show more considerable damage than one year of economic growth. Negative surprises are more likely than positive surprises of similar magnitude. Feedbacks that accelerate climate change are more prevalent than feedbacks that dampen warming, and the impacts of climate change are more than linear in climate change. Figure \ref{fig:totalimpact} illustrates this: The most pessimistic estimate is twice as large as the most optimistic one.

Estimates are not only uncertain but incomplete too. Some impacts\textemdash on violent conflict for example\textemdash are omitted altogether because they resist quantification. Other impacts are dropped because they do not fit the method\textemdash higher-order impacts in the enumerative method, non-market impacts in computable general equilibrium models. Assumptions about adaptation are stylised, either overly optimistic\textemdash rational agents with perfect expectations in markets without distortions\textemdash or overly pessimistic\textemdash dumb farmers doggedly repeating the actions of their forebears. Valuation of non-market impacts is problematic too as benefit transfer, the extrapolation of observed (or rather inferred) values to unobserved situations, has proven difficult \citep{Brouwer2000} yet is key to predicting how future people would value risks to health and nature. Comparing the sectoral coverage of various estimates, \citet{Tol2022meta} finds an average underestimate of 63\%.

The benefits of climate policy are the avoided impacts of climate change. The impact function described above predicts economic damages for alternative temperature trajectories, with and without climate policy, or with different intensities of climate policy. The difference between those impact trajectories constitute the estimate of the benefit of climate policy.

\section{A cost-benefit analysis}
\label{sc:cba}
Section \ref{sc:costs} reviews the costs of greenhouse gas emission reduction, Section \ref{sc:benefits} its benefits. I here put the two together in a cursory benefit-cost analysis of the temperature targets of the Paris Agreement.

Temperature trajectories are build with the carbon cycle model of \citet{MaierReimer1987} and the climate model of \citet{Schneider1981}, as parameterized in \citet{Tol2019bk}. In the baseline scenario the global annual mean surface air temperature reaches 4.8\celsius{} in 2100. This is hot, too hot probably, and so overestimates the benefits of climate policy.

\begin{figure}[h]
\includegraphics[width=\textwidth]{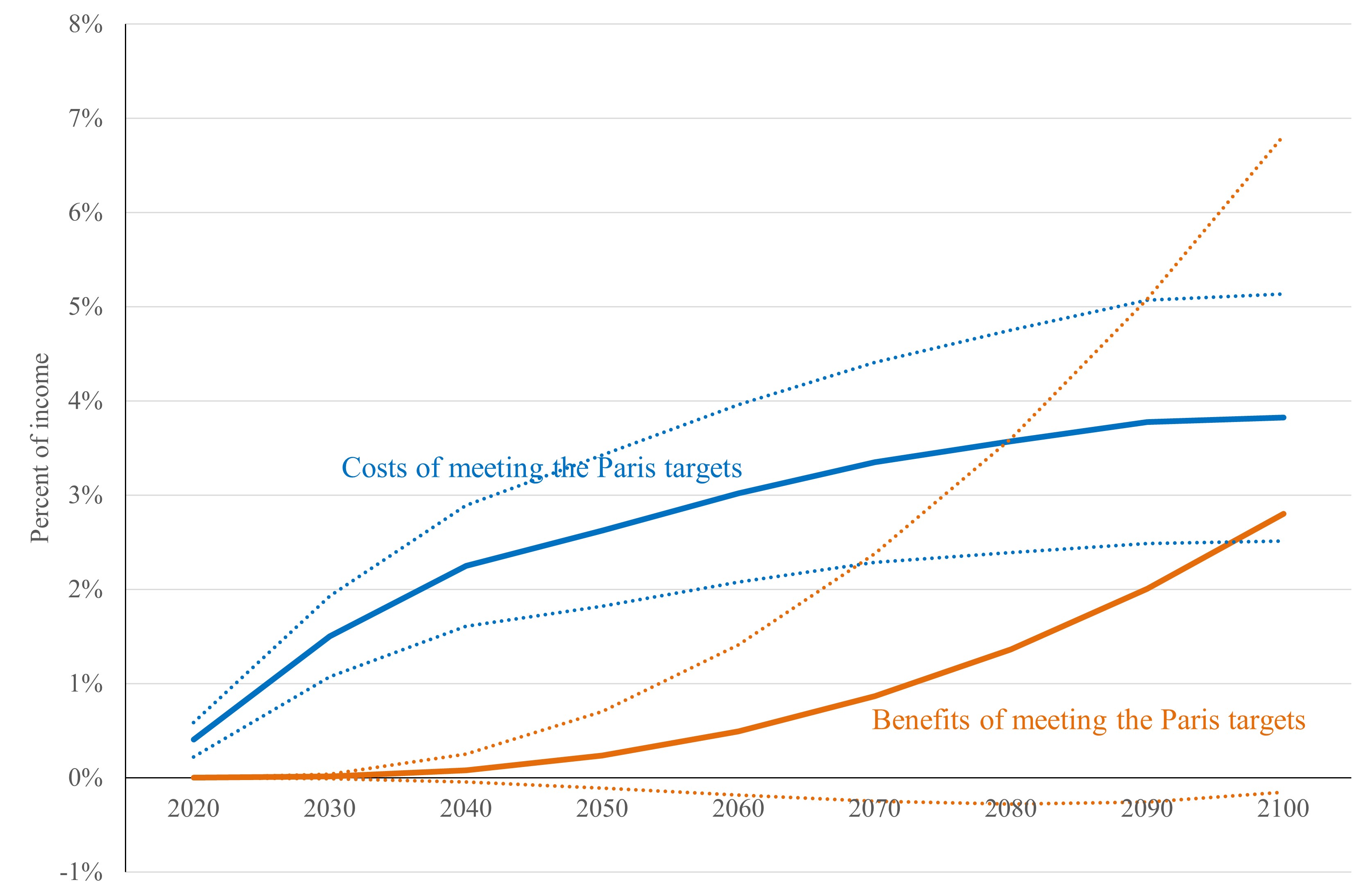}
\includegraphics[width=\textwidth]{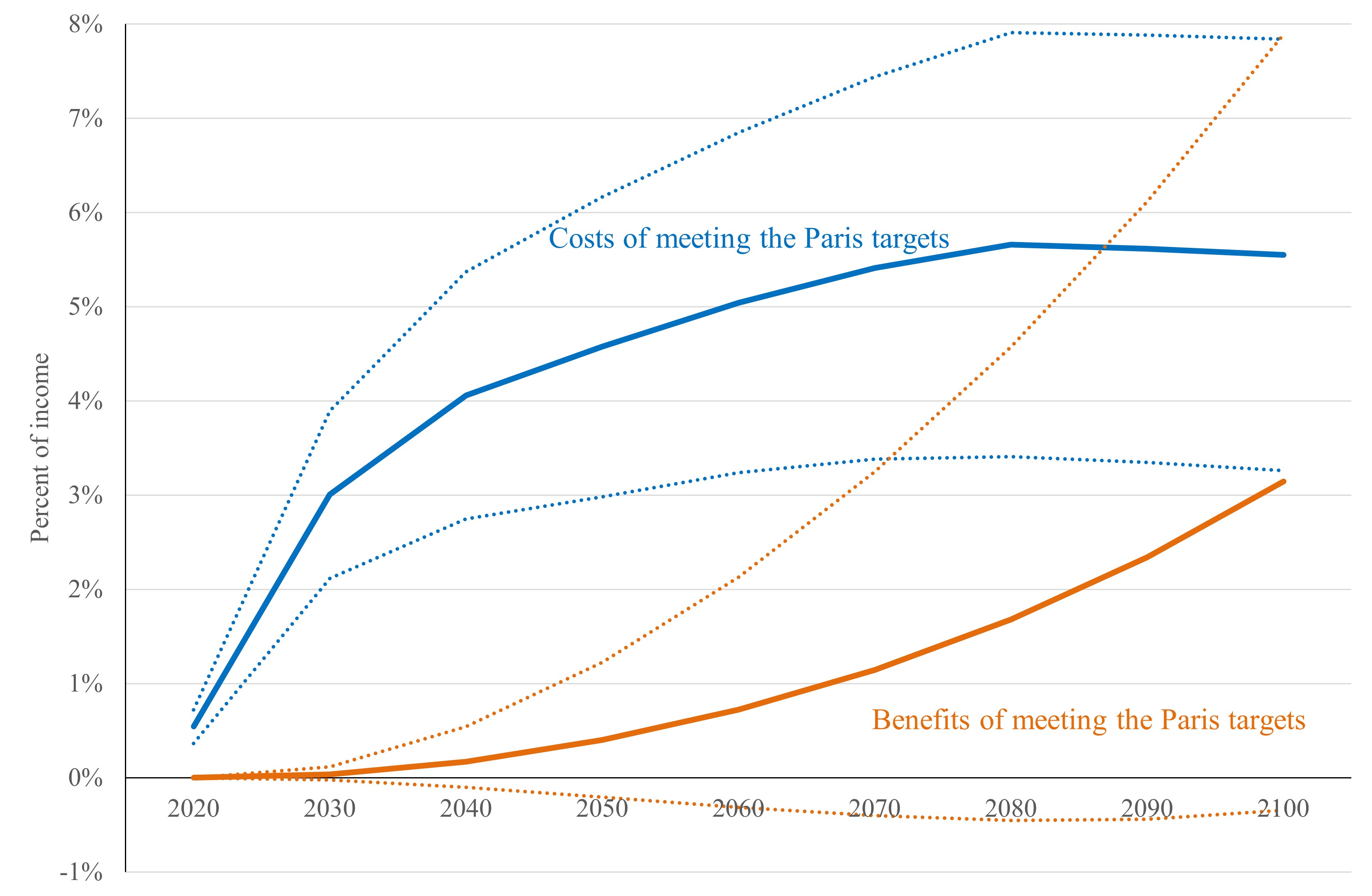}
\caption{Costs and benefits of meeting the Paris targets of 2.0\celsius{} (top panel) and 1.5\celsius{} (bottom) global warming.}
\label{fig:cba}
\end{figure}

Figure \ref{fig:cba} summarizes the key findings. The top (bottom) panel shows the costs and benefits of meeting the 2\celsius{} (1.5\celsius) target. The costs of the less ambitious target are just below 4\% of GDP in 2100, rising to just above 5.5\% of GDP for the more ambitious target. This is the average across models and scenarios in the IPCC 1.5\celsius{} Special Report database \citep{Rogelj2018IPCC}. The range shown is plus and minus the standard error across models.

Recall that these results assume first-best policy implementation. Even simple policy imperfections, such as a failure to equate carbon prices between countries, would readily double the costs of climate policy \citep[e.g.,][]{Boehringer2009}.

Figure \ref{fig:cba} also shows the benefits, here defined as the difference between the SSP5-8.5 scenario and the respective policy scenarios. The baseline scenario is unrealistically hot \citep{Srikrishnan2022}, which strengthens the case for emission reduction. Nevertheless, the benefits of climate policy are smaller than its costs, some 2.8\% of GDP for the 2\celsius{} target and about 3.1\% for 1.5\celsius. The range shown is again plus or minus what may be considered a standard error \citep[see][for its derivation]{Tol2022meta}.

If I instead use the SSP3-7.0 scenario as the baseline, the world would warm not 4.8\celsius but 3.9\celsius by 2100. The benefits of climate policy would then be 1.8\% of GDP for the 2\celsius{} target and 2.2\% for the 1.5\celsius{} target.

The central estimate of the benefits is always smaller than the central estimate of the costs. Ignoring the uncertainty for the moment, regardless of the discount rate, the present costs exceed the present benefits; the net present benefits are negative.

\begin{figure}[h]
\includegraphics[width=\textwidth]{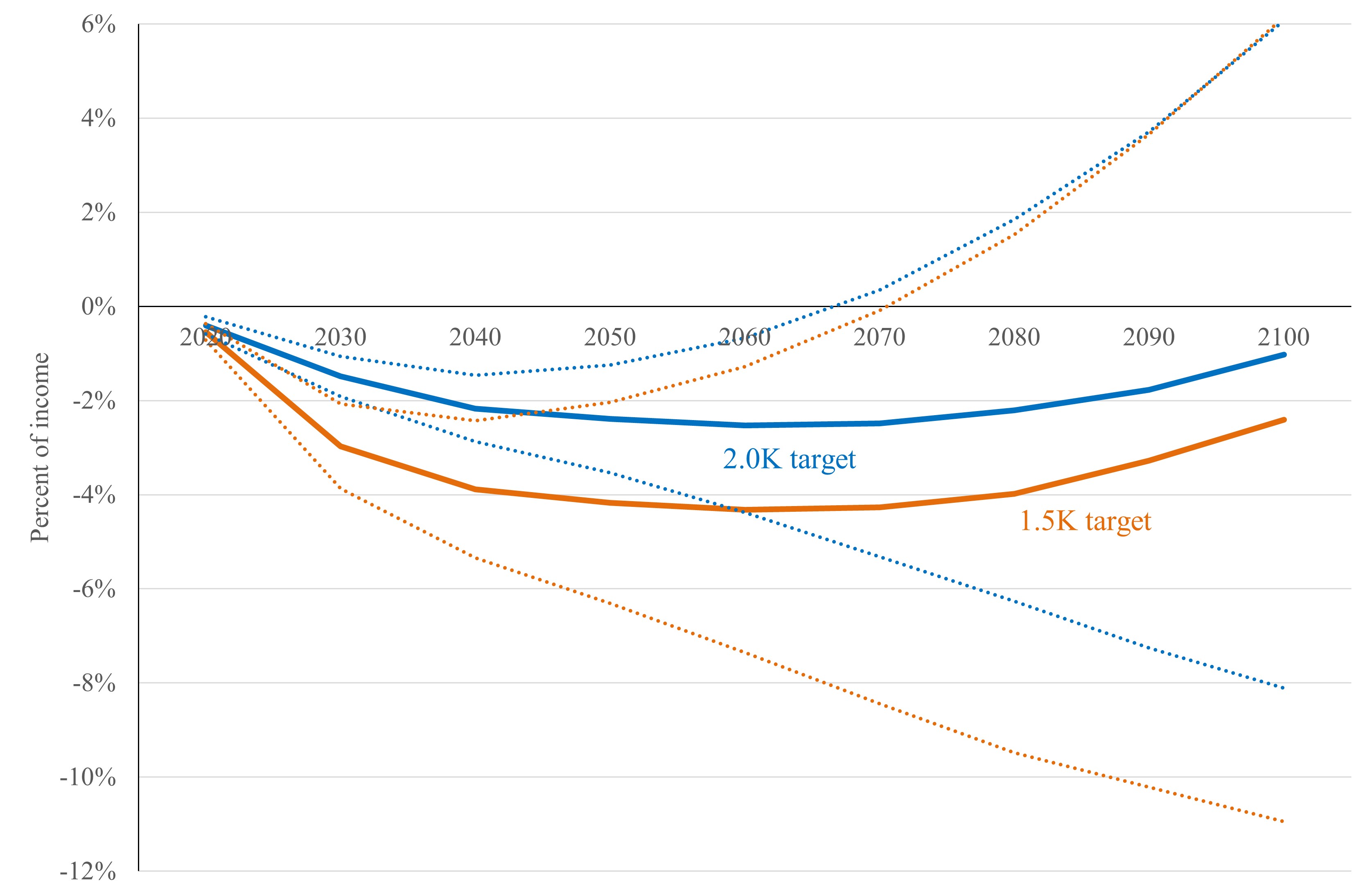}
\caption{Net benefits of meeting the Paris targets of 2.0\celsius{}  and 1.5\celsius{} global warming.}
\label{fig:netben}
\end{figure}

Figure \ref{fig:netben} shows the net benefits. It reaffirms that the central estimate of the costs is larger than the central estimate of the benefits. The central estimate is always negative. The confidence interval is rather large, however. From 2070 onwards, net benefits cannot be excluded. Without conducting a formal benefit-costs analysis, this confirms what is known from the literature: Stringent climate policy can be justified with a high rate of risk aversion and a low discount rate.

\section{Discussion and conclusion}
\label{sc:conclude}
This paper reviews the costs and benefits of climate policy and assesses the economic justification of the climate targets in the Paris Agreement. Assuming first-best policy implementation and the deployment of negative emission technologies yet to be demonstrated at scale, meeting the 2.0\celsius{} (1.5\celsius) target would cost just under 4.0\% (over 5.5\%) of GDP in the year 2100, with a considerable range of uncertainty. The benefits of these climate policies are smaller, just under (over) 3.0\% of GDP in 2100, but the uncertainty about the benefits is considerable larger than the uncertainty about the costs. The central estimate is that the costs exceed the benefits throughout the 21st century, but from 2070 onward net benefits cannot be excluded. Note that the above benefits of climate policy are inflated by the choice of an unrealistically warm baseline scenario, and its costs deflated by the use of first-best policy implementation. The Paris climate targets therefore only pass the cost-benefit test if the discount rate is low and the rate of risk aversion high.

The main finding is not new, indeed has withstood the test of time. Instead of trying to refine cost-benefit analyses of climate policy, research should therefore focus elsewhere. The number of \textit{ex post} estimates of the costs and efficacy of climate policy is growing rapidly. These cannot replace \textit{ex ante} studies, but should inform model parameterizations and perhaps encourage retirements too. The number of empirical studies of the impact of climate change is also growing rapidly. This information needs to be consolidated and absorbed into Integrated Assessment Models.

The biggest policy challenge lies in dealing with the inevitable fall-out if the 1.5\celsius{} target is missed, perhaps later this decade, and the 2.0\celsius{} becomes undeniably infeasible. The environmental movement will have to come to the terms with a catastrophe that was foretold but did not materialize. These topics are perhaps better left to political scientists and social psychologists.

Besides new and presumably better numbers in cost-benefit analysis, economists should focus on evaluating the many policy initiatives to reduce greenhouse gas emissions and the less numerous attempts to reduce vulnerability to climate change, as well as on the drivers of emissions and vulnerability that have little or nothing to do with climate policy.

\bibliography{master}

\begin{table}[h]
    \centering
    \caption{Carbon dioxide emissions from energy use and the four components of the Kaya Identity for four periods.}
    \label{tab:kaya}
\begin{tabular}{l r r r} \\ \hline
Indicator & \multicolumn{3}{c}{period} \\ 
& 1965-1999 &	1999-2011 & 2011-2021 \\ \hline
Population & 1.8\% & 1.2\% & 1.1\% \\
Income per capita & 1.7\% &	1.9\% & 1.5\% \\
Energy intensity & -0.8\% & -0.6\% & -1.1\% \\
Carbon intensity & -0.4\% & 0.4\% & -0.8\% \\
Emissions & 2.3\% & 2.9\% & 0.7\% \\ \hline
\end{tabular}
\end{table}

\end{document}